\documentclass[a4paper,twoside]{article}
\newcommand{\affil}[1]{$^{\rm #1}$}
\usepackage[authoryear]{natbib}
\bibpunct
{(}{)}{;}{a}{}{,}
\usepackage{graphicx}
\date{} 
\title{\large\bf\flushleft Multivariate analysis of the globular clusters in M87}
\author{\parbox{\textwidth}{\flushleft
\vspace{-0.5cm}
{\it Sukanta Das\affil{A}, Tanuka Chattopadhayay  \affil{A} and Emmanuel Davoust\affil{B}}\\
\vspace{0.4cm}
{\small \affil{A}\,Department of Applied Mathematics, University of Calcutta, 92 A.P.C Road, Calcutta 700009}\\
{\small \affil{A}\,Email: tanuka@iucaa.ernet.in} {\small
\affil{B}\,Institut de Recherche en Astrophysique et
Plan\'etologie, Universit\'e de Toulouse/CNRS, 14 Avenue Edouard
Belin, 31400 Toulouse, France }\\
{\small \affil{B}\,Email: edavoust@irap.omp.eu}}}
\begin{document}
\twocolumn[
\begin{changemargin}{.8cm}{.5cm}
\begin{minipage}{.9\textwidth}
\vspace{-1cm} \maketitle
%
%
\small{\bf Abstract:} \noindent An objective classification of 147
globular clusters in the inner region of the giant elliptical
galaxy M87 is carried out with the help of two methods of
multivariate analysis. First independent component analysis is
used to determine a set of independent variables that are linear
combinations of various observed parameters (mostly Lick indices)
of the globular clusters. Next K-means cluster analysis is applied
on the independent components, to find the optimum number of
homogeneous groups having an underlying structure. The properties
of the four groups of globular clusters thus uncovered are used to
explain the formation mechanism of the host galaxy. It is
suggested that M87 formed in two successive phases. First a
monolithic collapse, which gave rise to an inner group of
metal-rich clusters with little systematic rotation and an outer
group of metal-poor clusters in eccentric orbits. In a second
phase, the galaxy accreted low-mass satellites in a
dissipationless fashion, from the gas of which the two other
groups of globular clusters formed. Evidence is given for a blue
stellar population in the more metal rich clusters, which we
interpret by Helium enrichment. Finally, it is found that the
clusters of M87 differ in some of their chemical properties (NaD,
TiO1, light element abundances) from globular clusters in our
Galaxy and M31.


\medskip
\medskip
\end{minipage}
\end{changemargin}
] \small

\section{Introduction}

\noindent  The details of galaxy formation and evolution remain
one of the great unsolved problem in modern astrophysics. Giant
elliptical galaxies at the centre of galaxy clusters are of strong
interest as they provide case studies for galaxy formation theory.
On the other hand globular clusters (GCs) are excellent tracers of
galaxy halos. Indeed some 50\% of the stars in our Galaxy may have
originated in GCs (Martell \& Grebel 2010). GCs are known to form
in vigorous star formation events (Brodie \& Strader 2006 and
references therein) and were an important mode of star formation
in the early universe (Muratov \& Gredin 2010). It is possible
that all stars may have originally formed in star clusters which
are only later dispersed into the smooth stellar halos we see
today (Lada \& Lada 2003). For making progress in galaxy formation
theory, a holistic approach is required, which involves
compositional as well as kinematical properties of GCs in giant
elliptical galaxies. A landmark study (Emsellem et al. 2007)
suggested that a fundamental dichotomy exists between dynamically
'slow' and 'fast' rotating elliptical galaxies. These observations
prompted new interpretations and theoretical models to explain
this difference (Krajnovic et al. 2008; Jesseit et al. 2009).
Proctor et al. (2009) suggested that the rotator class may
change when the kinematics are probed beyond the inner regions.\\

\noindent According to various studies, classical formation of
galaxies have been proposed along five major lines: (i) the
monolithic collapse model, (ii) the major merger model, (iii) the
multiphase dissipational collapse model, (iv) the dissipationless
merger model and (v) accretion and in situ hierarchical merging.
But no model is globally
acceptable over others.\\

\noindent  In this context, one is tempted to make a detailed
study of an archetype of each category of elliptical galaxy. In
previous papers (Chattopadhyay et al. 2009; Chattopadhyay et al.
2013b) we studied the GCs of NGC5128, which is a slowly rotating
elliptical galaxy. In the present paper we study the GCs of M87,
which is a strongly rotating elliptical galaxy, located at the
centre of the Virgo cluster. It has a large population of GCs with
well-known kinematic (Huchra \& Brodie 1987; Cohen \& Ryzhov 1997;
Hanes et al 2001, C\^ot\'e et al. 2001; Strader et al. 2011),
photometric (Strom et al. 1981; Strader et al. 2011) and chemical
properties (Mould et al. 1987, 1990; Cohen et al. 1998). To this
end, we carry out a multivariate analysis of the Lick indices,
metallicities and radial distances of the GCs. First we use
independent component analysis (hereafter ICA), suitable for
non-Gaussian data sets to search for the independent components
(hereafter IC). The number of such components, which are linear
combinations of parameters, is equal to the number of parameters
(p, say). Since only a few (say, m$<<p$) of the IC components may
explain a larger percentage of variation in the data, one can take
only those m components instead of all p components. Then the GCs
are classified on the basis of those m independent components
using another exploratory data analytic method, namely K-means
cluster analysis (Chattopadhyay et al. 2009, 2010, 2012, 2013a,b)
to find the homogeneous groups. In the end, the properties of the
homogeneous groups of GCs allow us to propose a possible scenario
for the formation of
the GCs and their host galaxy.\\

\noindent The data sets used are presented in section 2. Section 3
describes different methods used in the study. The results and
interpretations have been included in Section 4.

\section{Data set}

\noindent The Lick indices ($H\beta$, Mg1, Mg2, Mgb, Fe5270,
Fe5335, NaD and TiO1)  and metallicities (Fe/H) used in the
present analysis were taken from Cohen et al. (1998). From the
original sample of 150 GCs we removed the GCs with identification
numbers 5024 and 5026, because they are redundant with 978 (Hanes
et al. 2001), as well as 321 which is a star (Strader et al.
2011), thus reducing the sample to 147 GCs. In the latter paper
the identification number given by Cohen et al. (1998) is preceded
by "S". The velocities $v_{rad}$, half-light radii $r_h$, $i$
magnitudes ($i_{mag}$) and $(g - r)$ and $(g - i)$ colours were
taken from Strader et al. (2011). The radial distances $R$ and
position angles $\psi$ were derived from the coordinates listed in
Strader et al. (2011), assuming that the centre of of the system
of GCs is also that of the galaxy, which is at $\alpha_c$ = 12h
30m 49.42s
and $\delta_c$ = 12d 23' 28.044" (Lambert \& Gontier 2009).\\

\noindent It is to be noted that Our sample size is small compared
to the recently observed samples by various authors (e.g. Strader
et al. 2011). We have thus plotted the histograms of i -
magnitudes and radial distances of our sample and have compared
with the sample used by Strader et al. (2011) (Fig.1). Now Strader
et al.(2011) have found that the bright and faint GCs have the
same velocity dispersions (sigma = 340 and 339 km /s). In their
section 6.3.4, they find that there is no strong trend with
magnitude for the red GCs. For the blue GCs there is perhaps an
avoidance of the systematic velocity, although marginal. The
conclusion is that the bright and faint compact blue GCs belong to
the same population. So we can safely say that, if the kinematics
are an indicator of the origin of GCs, then our sample
should be enough to determine the origin of the inner GCs.\\

\noindent For comparison purposes, we used the Lick indices of 313
old clusters in M31 (Schiavon et al. 2012), those of 47 GCs in
4472 (Cohen et al. 2003) and 33 GCs in NGC 4636 (Park et al.
2012).

\section{Statistical methods}

\subsection{Shapiro-Wilk test}

\noindent The choice of statistical methods to apply to a set of
data depends on its Gaussian or non Gaussian nature. We have thus
tested the Gaussianity of our data set by the Shapiro-Wilk test
(1965). According to this test the null hypothesis is that the
data set is Gaussian. The test statistic W is defined as
W=$\Sigma_{i=1}^n a_i
x_i^2$/$\Sigma_{i=1}^n (x_i -\bar{x})^2$\\

\noindent where n is the number of observations, $x_i$ are the
ordered sample values and $a_i$ are constants generated from
means, variances and covariances of the order statistics of a
sample of size n from a normal distribution. In our case the data
set is multivariate, hence we have used the multivariate extension
of the Shapiro-Wilk test. We found that the p value of the test is
$1.327 \times 10^{-11}$, which is very small. Thus the null
hypothesis is rejected at a 5\% level of significance. We conclude
that the
present data set is non Gaussian in nature.\\

\subsection{Independent component analysis}

\noindent While Principal Component Analysis (PCA) applies to
Gaussian data sets, ICA applies to non Gaussian data sets. ICA is
a dimension reduction technique, i.e. it reduces the number of
observable parameters p to a number m (m$<<$p) of new parameters,
where these new parameters are the linear combinations of p
parameters, such that these m parameters are mutually independent.
Mathematically speaking let $X_1,X_2,......,X_p$ be p random
vectors (here p parameters) and n (here 147) be the number of
observations of each
$X_i$ (i=1,2,...,p).\\

Let
\begin{equation}
X = AS
\end{equation}
\noindent where $S = [S_1, S_2, ......, S_p]^{'}$ is a random
vector of hidden components $S_i$ (i=1,2,...,n) such that $S_i$
are mutually independent and A is a non singular matrix. The goal
of ICA is to estimate A and to find S by inverting A i.e.
\begin{equation}
S = A^{-1} X \ \ i.e. \  \   S = WX
\end{equation}

\noindent  Under ICA we find the components $[ S_1, S_2, ......,
S_n]$ those are independent by finding an unmixing matrix W in
such a way that the covariance between any two nonlinear functions
$g_1(S_i)$ and $g_2(S_j)$ for $ i \ne j $ is zero i.e. the
independent components are nonlinearly uncorrelated (for more
details see Comon (1994); Chattopadhyay et al. (2013b) and
references
therein). \\

\noindent At present there is no better method available to
automatically determine the optimum number of Independent
Components (ICs). In this paper, the number of ICs is determined
by the number of Principal Components (PCs) chosen (Albazzaz and
Wang, 2004). To reduce the number of components $S_i$ from p to m
(m$<<$p), one is to perform PCA (Babu et al. 2009; Chattopadhyay
\& Chattopadhyay 2007; Fraix Burnet et al. 2010; Chattopadhyay et
al. 2010). In this method also $Y_i$'s (i=1,2,...,p) vectors are
found, which are linear combinations of $X_i$'s (i=1,2,...,p) such
that $Y_i$'s are uncorrelated. The number of (initially p)
components is reduced to m by taking those $Y_i$'s (i=1,2,...,m)
for which the corresponding eigenvalues $\lambda_i \sim 1$.\\

\noindent  In the present work we have first performed PCA to find
the significant number of IC components to be chosen. PCA applies
to Gaussian data, but the present data set is non Gaussian, so we
performed ICA. In PCA the maximum variation  with significantly
high eigenvalue (viz.$\lambda \sim 1$) was found to be almost 78\%
for three PCA components. Hence we have chosen three IC components
for the cluster analysis. The parameters chosen for ICA are the
Lick indices $H\beta$, Mg1, Mg2, Mgb, Fe5270, Fe5335, NaD, TiO,
the metallicity Fe/H, the photometric parameters magnitude
$i_{mag}$ and colours (g-r), (g-i) and the projected
galactocentric distance $R$ (in arcsec).

\subsection{K-means cluster analysis}

K-means cluster analysis (CA) is a multivariate technique for
finding homogeneous groups in a data set giving information of the
underlying structure in this data set. In this method one finds k
groups, provided each group contains an object and each object
belongs to exactly one group. Hence the maximum and minimum number
of possible groups are k = n and k = 1 respectively. The algorithm
is as follows :
\begin{itemize}
\item (i) All the objects are divided into k (given) groups in a
random manner. \item (ii) Any group is selected and any particular
object of that group is taken first. Then the parametric distance
(here the parameters are 3 IC components) from the chosen object
is computed for the remaining objects. If the distance between
objects in the group is greater than that for objects in other
groups, they are interchanged. \item (iii) Process (ii) is applied
for all objects and for all groups. \item (iv) Steps (i) and (iii)
are continued until there is no further change.
\end{itemize}
The k groups thus found are coherent in nature.\\

\noindent  The optimum value of k is found as follows. First one
finds groups for k = 1, 2, 3,...etc. Then a distance measure $d_k$
is computed by
\begin{equation}
d_k=(1/p)min_xE[(x_k-c_k)'(x_k-c_k)]
\end{equation}
\noindent which is the distance of $x_k$ vector (values of
parameters) from the centroid $c_k$ of the corresponding group.
The optimum value of k is that for which $J_k = (d_k^{'
-p/2}-d_{k-1}^{' -p/2})$, is maximum (Sugar \& James 2003). In the
present case k = 4.\\

\noindent  For cluster analysis (CA) the number of parameters are
3 IC components chosen as previously described. The optimum number
of groups is selected objectively by k-means clustering (Mac Queen
1967), together with the method developed by Sugar and James
(2003). The optimum number is k = 4. The groups are labelled G1,
G2, G3 and G4.\\

\noindent For testing the robustness of the classification we have
proceeded in the following manner. In the original scheme the
number of parameters is 13. First we have constructed the variance
covariance matrix of the parameters and selected the parameters
having maximum variances e.g. here $H_{\beta}$, Mgb, Fe5270,
Fe5335, NaD, Fe/H, $i_{mag}$, R  have variances greater than 0.25
and the remaining ones, Mg1, Mg2, TiO1, g-r, g-i have variances
$\sim$0.0. Therefore we have done the analysis with the above 8
parameters and have obtained exactly the same groups with no
variation at all. So we can at once say that our classification is
robust with respect to the parameters which are responsible for
the maximum
variation.\\

\noindent Regarding the uncertainties on the various parameters,
the method we have used is totally exploratory (i.e. no underlying
distribution is assumed). Hence it is not possible to see the
effect of error in such type of analysis directly. What we can  do
at most is to change the values of the parameters within the range
permitted by the error bars and redo the analysis. We have tested
it for that for few values of the parameters and found the same
groups. Alternatively once the optimum classification (clustering)
is obtained, one can use a process called discriminant analysis
(Johnson \& Wichern 1998) to verify the acceptability of the
classification of different GCs. In this standard procedure, using
the probability density functions in parameter space for the
different clusters, one can assign an object (in this case a
globular cluster) to be a member of a certain class. If the
original classification is robust, then every GC should be
classified again as a member of the same class that it was before.
If a significant number of objects are not reclassified then that
means that the original classification is not stable and hence not
trustworthy. Table 1 shows the result of a discriminant analysis,
where the columns represent how the GCs of a cluster were assigned
by the analysis. The fraction of correct classifications is 0.925,
which implies that the classification is indeed robust by
computing classification /misclassification probabilities for the
GCs.

\subsection{Levenberg-Marquardt Algorithm}

\noindent This algorithm is used to find the most probable mean
rotation curve for each group, assuming that they have a
solid-body rotation around the centre of the galaxy. This is a
reasonable assumption for the two innermost groups, G1 and G4,
because the galaxy itself is in solid-body rotation at least to a
radius of 225" (Cohen \& Ryzhov, 1997). For the two other groups,
G2 and G3, we also tried a rotation curve that flattens beyond
225", but the data were better adjusted
by a solid-body rotation curve (also see Fig.7 of Kissler-Patig \& Gebhardt 1998).\\

\noindent The rotation amplitudes $v_{rot}$ ( = $\Omega R$) and
the position angles ($\psi_0$) of the axes of rotation (East to
North) of the different groups G1 - G4, (found in the CA) are
computed by the relation

\begin{equation}
v_{rad}(\psi)=v_{sys}+\Omega R\sin(\psi-\psi_0)
\end{equation}

\noindent (C\^ot\'e et al. 2001; Richtler et al. 2004; Woodley et
al. 2007), where $v_r$ is the observed radial velocity, $v_{sys}$
is taken as the recession velocity of the galaxy, which is 1287
km/s.\footnote{C\^ot\'e et al. (2001) adopted $v_{sys}$=1350
km/s.}, $R$ is the projected distance of each GC and $\psi$ is its
position angle, measured East to North. We have used
Levenberg-Marquardt fitting method (Levenberg 1944; Marquardt
1963) to solve for $v_{rot}$ and $\psi_0$. They are listed in
Table 2 for G1 - G4. Now kinematic data sets for G1 - G4 are small
and for such situation  Monte Carlo simulations are used to
increase the data sets to have a more convincing result. Monte
Carlo simulation needs distributional assumption and here no well
known bivariate distribution is fitting with the data well. So we
have taken several bootstrap samples and have computed the mean
values with standard errors of the rotation parameters . The
errors are small as seen from Table 2.

\section{Results and interpretation}

\subsection{Properties of globular clusters in the four groups}

\noindent The cluster analysis divided the GCs of M87 into four
groups. We did not use the kinematical data ($v_{rad}$ and $\psi$)
in the statistical analysis. Nor did we use  Mgb/Fe. The latter is
an indicator of the abundance in light elements. But we did use
these parameters for interpreting the results.\\

\noindent The results of the analysis are summarized in Table 2,
which lists the mean values of all the parameters with standard
errors for the four groups. The mean velocities of rotation and
mean position
angles have been computed by the Levenberg-Marquardt algorithm.\\

\noindent To show how the groups are separated we have plotted
three Independent Components (viz. IC1, IC2 and IC3) for the
groups G1 - G4 found in our analysis (Fig.2). That figure shows
that the groups are well separated in IC space.\\

\noindent  We first investigate how the different groups differ in
their Lick index values, and how they compare in this respect with
the simple stellar population models of Thomas et al. (2011). As
shown on Figs.3, 4 and 5, G3 has the lowest values of Fe5270,
NaD,TiO and Mgb, while G4 has the highest values of these Lick
indices, and G1 and G2 are in between. These two latter groups
differ in that G1 is marginally less chemically evolved than G2
(these two groups also differ in their spatial distribution and in
their kinematics). In other words, the order of increasing overall
chemical evolution is G3, G1, G2, G4. The Lick index H$\beta$ is
an age indicator, higher values meaning younger ages, but this
index is also sensitive to the colour of the Horizontal Branch
(HB), higher values meaning a bluer HB. A bluer HB in turn can be
due to He-enriched stars. So the interpretation of Fig.6, which
shows the run of H$\beta$ vs Mgb, is best done with the help of
stellar population models, namely those of Thomas et al. (2011).
The data are well fit by models of 12Gyr, albeit with a large
scatter suggesting a variety of HBs and/or of $\alpha$/Fe. An
alternative explanation in terms of stellar populations with
younger ages is not possible because it would not be compatible
with the observed broad-band colours (see below).\\

\noindent We next investigate the colours of the stellar
populations in the GCs of M87. The colour-colour diagram for the
different groups is shown in Fig.7. There is a progressive
reddening of the populations, from G3 (green) to G1 (blue)
followed by G2 (red) and G4 (black). This is predicted by models
of synthetic stellar populations as a result of increasing age
and/or metallicity. Since the models of Thomas et al. (2011) do
not predict the evolution of broad-band colours, we adjusted
Yonsei models of simple stellar populations (Chung et al. 2013) to
the data. The models for metallicities ranging from [Fe/H] = -2.6
to 0.6, an age of 12 Gyr and $\alpha/Fe$ = 0.3 are shown as solid
lines, green and blue for 0 and 70\% secondary populations
respectively. Changing $\alpha/Fe$ will only shift the model down
diagonally on the figure, and changing the initial mass function
has little effect on the models (see Chung et al. 2013). The
comparison of our data to the Galev models (Kotulla et al. 2009),
and to the models of Maraston et al. (2003) gave results (not
shown) very close to those of the Yonsei model for 0\% secondary
populations. For the Maraston models, we converted the CFH-Megcam
magnitudes to
SDSS using the paper by Betoule et al. (2013).\\

\noindent The colours of the metal-poor GCs (shown on Fig.7) are
well fit by the models. On the other hand, the metal-intermediate
(G1) and metal-rich (G2) GCs can only be explained by a
Helium-enriched secondary stellar population (Chung et al. 2013).
No Yonsei model population with a realistic age is able to account
for the colours of group G4. We have no explanation for the colors
of this group, other than that they are unusual. It is not
possible to advocate younger stellar populations instead of
He-enriched ones to interpret the colors of the groups, because
such populations are bluer in both colours, but more so in (g -i).
Further evidence for He enhancement in the metal-intermediate and
metal-rich GCs of M87 is the far ultraviolet excess detected in
the GCs of M87 by Sohn et al. (2006), an excess which increases
with metallicity. Kaviraj et al. (2007) interpret this excess by
the possible presence of hot HB stars from super-He-rich stellar
populations, even though
their predicted ages are unrealistically old.\\

\noindent We next compare the properties of the GCs of M87 with
those of other galaxies. The GCs of M87 have values of Fe5270 and
H$\beta$ at a given Mgb (see Figs.3 and 6) that are comparable to
those of M31, but generally lower values of NaD and TiO at a given
Mgb (see Figs.4 and 5), except G4. In the latter two figures, the
predictions of the simple stellar population models seem to hold
the middle ground between the GCs of M87 and of M31. Concerning
NaD, the high values observed in M31 might be partly due to
interstellar absorption and thus less reliable. We are indebted to
the referee for this remark. To better understand the significance
of these lower values, we also plotted the same data for GCs of
two other elliptical galaxies in Virgo, NGC 4472 and 4636 on
Figs.4 and 5. The GCs of these two other ellipticals show the same
behaviour as those of M87, suggesting that GCs in elliptical
galaxies are different from those of spirals in this respect,
except at very high metallicities. We speculate that Na is
underabundant in G1-G3 and that the Na - O anti-correlation
present in Galactic GCs, interpreted as a result of multiple
stellar populations, is only present in G4.\\

\noindent The abundance of light elements tends to decrease with
increasing metallicity in the GCs of our Galaxy (Fraix-Burnet et
al. 2009), M31 (Colucci et al. 2012a) and the LMC (Colucci et al.
2012b), as expected from arguments of nucleosynthesis.  High
values of $\alpha$/Fe result from star formation that occurs
before type 1a supernovae significantly increase the abundance of
Fe in the interstellar medium, and are expected mainly in
metal-poor environments.  In M87, we find that the opposite is
true. We use Mgb/Fe = 2Mgb/(Fe5270 + Fe5335) as an indicator of
$\alpha$/Fe and calibrate this indicator with the stellar
population models of Thomas et al. (2011). As shown on Fig.9, the
light-element abundance increases with metallicity. For G1 and G2
it increases from 0 to 0.3, and for G4 it is about 0.4. For G3,
which is very metal-poor, the disagreement between $\alpha$/Fe and
its indicator Mgb/Fe suggests that the light-element ratio
estimates, which are more uncertain at low metallicity, are just
not reliable for this group, which is ignored. The contrast
between the GCs of M87 and those of our Galaxy and M31 in this
respect are further evidence that GCs of elliptical and spiral
galaxies have different chemical histories.

\noindent We have adjusted a mean rotation curve to each of the
four groups (see Sect.3.4). The resulting parameters are listed in
Table 2, and the mean rotation curves of the four groups along
with radial velocities of the GCs are shown in Fig.10. The only
groups with clear evidence of rotation are G2 and G4. The outer
(radius $>$ 300 arcsec) clusters of the group G2 rotate faster
than the inner ones. Schubert et al.(2010) have found that the
metal poor GCs in the outer region of NGC 1399 (the non-rotating
giant elliptical galaxy in the centre of the Fornax cluster) is
the only GC population that shows significant rotation. The
difference with M87 is that G2 is not the metal poorest group.
There is marginal evidence for rotation in G1, about an axis which
is orthogonal to that of G2. There is also evidence for rotation
in G4, whose spatial distribution
 is flattened in the plane of rotation (along position angles of maximum or minimum
radial velocity; see Fig.10). As G4 is centrally concentrated, it
is worth mentioning the recent study of Emsellem et al. (2014),
who find a counterrotating core in the central 2 arcsec of M87, in
position angle 20 degrees. In a detailed kinematic study of the
GCs of M87, Strader et al. (2011) have found subpopulations with
distinct kinematical behaviours, and concluded that M87 is still
in active assembly. They also warned that the old kinematical data
(which include the present ones) predict a higher rotation than
their new data. Our own kinematical results should thus be viewed
with caution.

\subsection{Origin of the four groups}
\noindent The first stage of formation of a galaxy is a
dissipative collapse. During that phase, chemical enrichment is
expected to be accompanied by spin up of the metal-richer material
which ends up in circular orbits, whereas the metal-poor material
is in eccentric orbits (Eggen et al. 1962). The formation of G3
and G4 can be associated to that phase : G3 in the outer parts
from metal-poor material, G4 in the inner part from metal-rich
material.  The lack of rotation in G3, the flattened spatial
distribution, strong rotation and high average $\alpha$/Fe in G4,
are properties consistent with this assumption. Montes et
al.(2014a) have shown that the central part of M87 is super solar
and presumably formed first during the monolithic collapse of the
galaxy.\\

\noindent Montes et al.(2014b) have found that the innermost GCs
of M87 (within 30 arcsec) are metal poorer than the stars in the
region. We show that there are actually two populations of GCs (G3
and G4) near the centre, one
of solar metallicity like the stars of the central region (Montes et al. 2014a).\\

\noindent In the second stage of formation, the galaxy accretes
low mass satellites in a dissipationless fashion (Bekki et al.
2005). The GCs of G1 and G2 may have formed in this phase. The
difference between G1 and G2 could be one of time, G2 forming
first and retaining the angular momentum of the gas from which it
formed and G1 forming later from cooling flows close to the
centre, the gas having lost its angular momentum to collision with
gas clouds (viz. Table2, $v_{rot}$).  This would explain why the
GCs of G1, near the centre, have little rotation, while those of
G2, in the outer regions, rotate significantly. In this scenario,
one thus expects the velocity of rotation to increase outward,
which seems to be the case for the GCs of G2 : $v_{rot}\sim$
374.93 km$s^{-1}$ for $R\geq$ 300, and $v_{rot}\sim$ 179.92
km$s^{-1}$) for $R<$ 300.

\section{Conclusion}

\noindent We have performed a multivariate analysis of the
chemical and photometric properties of the GCs of M87, and found
four groups of clusters with distinct properties. The latter are
used to propose a formation scenario for the different groups, in
terms of dissipational collapse of the galaxy, followed by
dissipationless accretion of matter.\\

\noindent Regardless of the clustering results, the reanalysis of
data from the literature with recent stellar population models,
those of Thomas et al. (2011) and the Yonsei models (Chung et al.
2013), provides new insights into the chemical properties of the
GCs of M87, compared to those of other galaxies. We have found
that these properties are different from those of GCs in M31, in
that the Lick indices NaD and TiO1 are higher at a given Mgb. We
have also found a progressive excess blueing of the GCs with
metallicity, which we interpret by the presence of an extreme blue
HB, possibly due to He enriched secondary stellar populations.
Finally, $\alpha$/Fe increases with metallicity, at variance with
what occurs in GCs of spiral galaxies (our own and M31). Our
interpretation is one among many others and remains speculative in
view of the smallness of the sample and is not an ambitious as is
needed. Our interpretation is only one possible  way of
understanding the results.

\clearpage

\begin{figure}[h]
\begin{center}
\includegraphics[scale=0.75, angle=0]{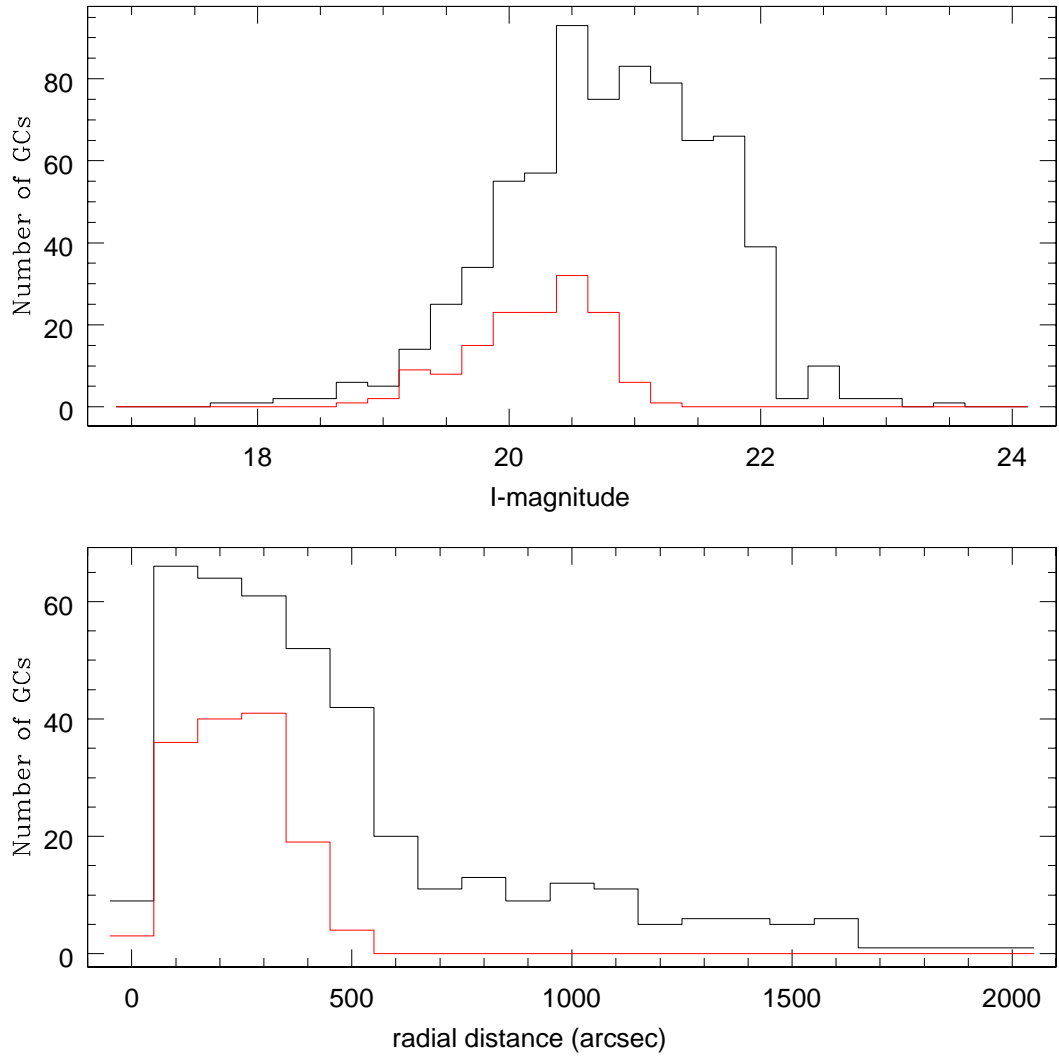}
\caption{Comparison of the histograms of $i_{mag}$ and R of the
GCs of our sample (red solid line) and the sample of Strader et
al. (2011) (black solid line)}\label{Fig.1}
\end{center}
\end{figure}
\clearpage

\begin{figure}[h]
\begin{center}
\includegraphics[scale=0.75, angle=0]{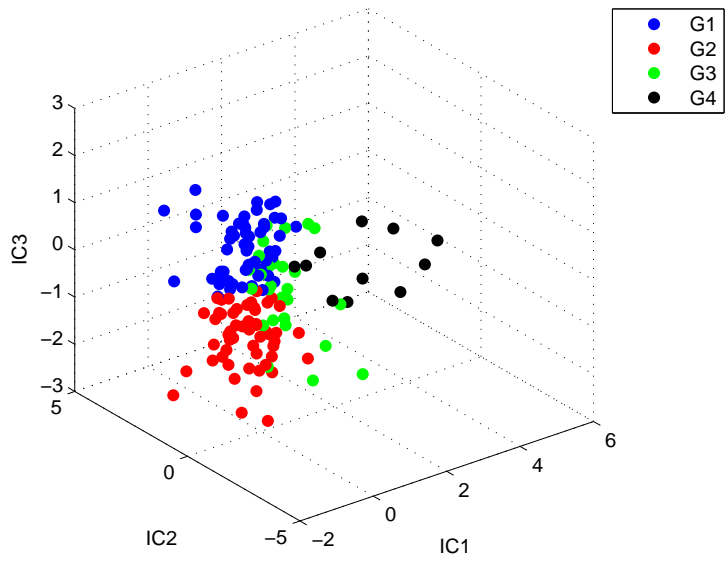}
\caption{Plot of IC1, IC2 and IC3 of G1(blue circles), G2(red
circles), G3(green circles) and G4 (black circles) groups of
GCs.}\label{Fig.2}
\end{center}
\end{figure}

\clearpage

\begin{figure}[h]
\begin{center}
\includegraphics[scale=0.75, angle=0]{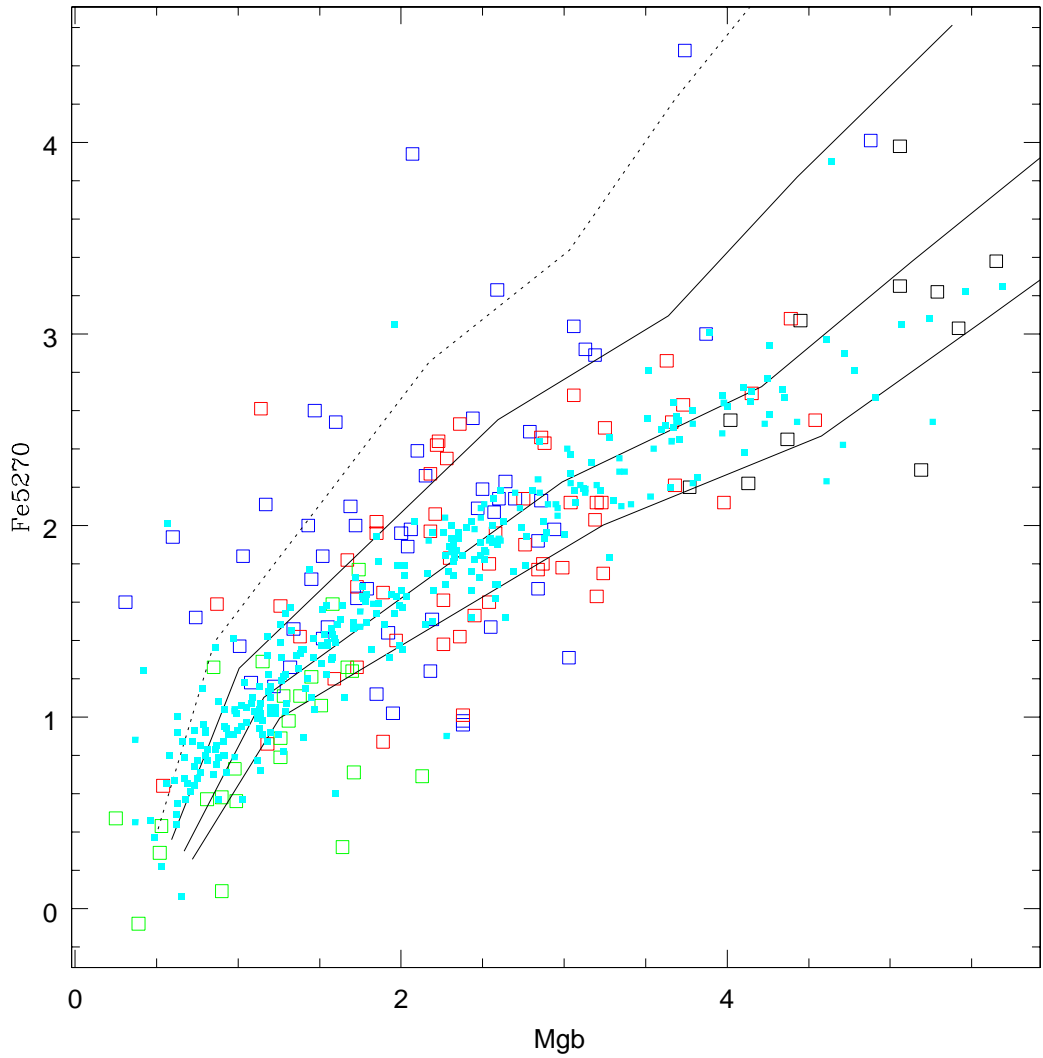}
\caption{Fe 5270 vs Mgb for G1 - G4. Blue open squares for G1, red
open squares for G2, green open squares for G3, black open squares
for G4, cyan filled squares for GCs of M31. Black lines are
stellar population models of Thomas et al.(2011) for an age of 12
Gyr, a dotted line for $\alpha/Fe$ = -0.3 and solid lines are for
for $\alpha$/Fe = 0.0, 0.3 and 0.5 respectively from left to
right.}\label{Fig.3}
\end{center}
\end{figure}

\clearpage

\begin{figure}[h]
\begin{center}
\includegraphics[scale=0.75, angle=0]{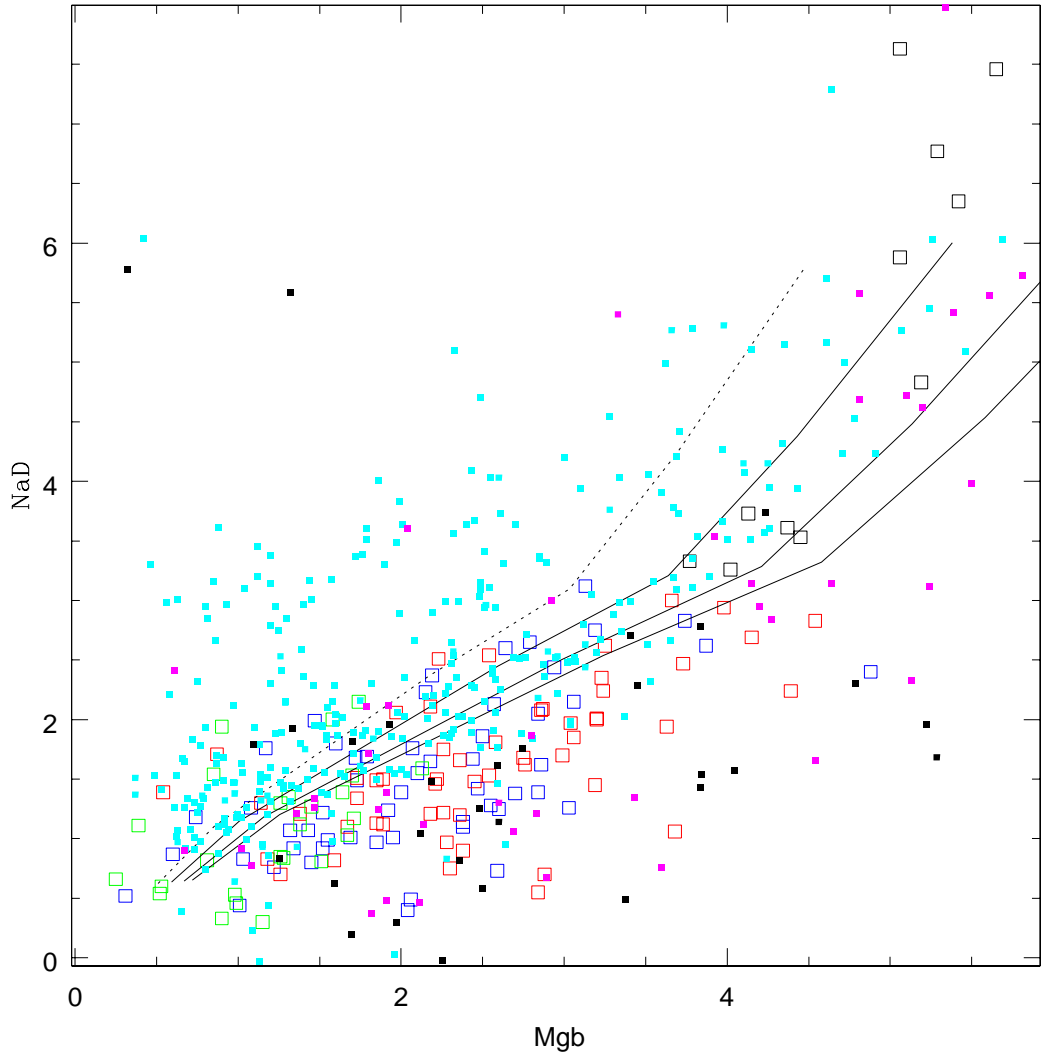}
\caption{NaD versus Mgb for G1 - G4 and for GCs of M31 (small cyan
squares), NGC4472 (small magenta squares), NGC4636 (small black
squares). The colours and symbols are the same as in Fig.3. NaD is
higher in the GCs of the spiral galaxy than in those of the
elliptical galaxies.} \label{Fig.4}
\end{center}
\end{figure}

\clearpage

\begin{figure}[h]
\begin{center}
\includegraphics[scale=0.75, angle=0]{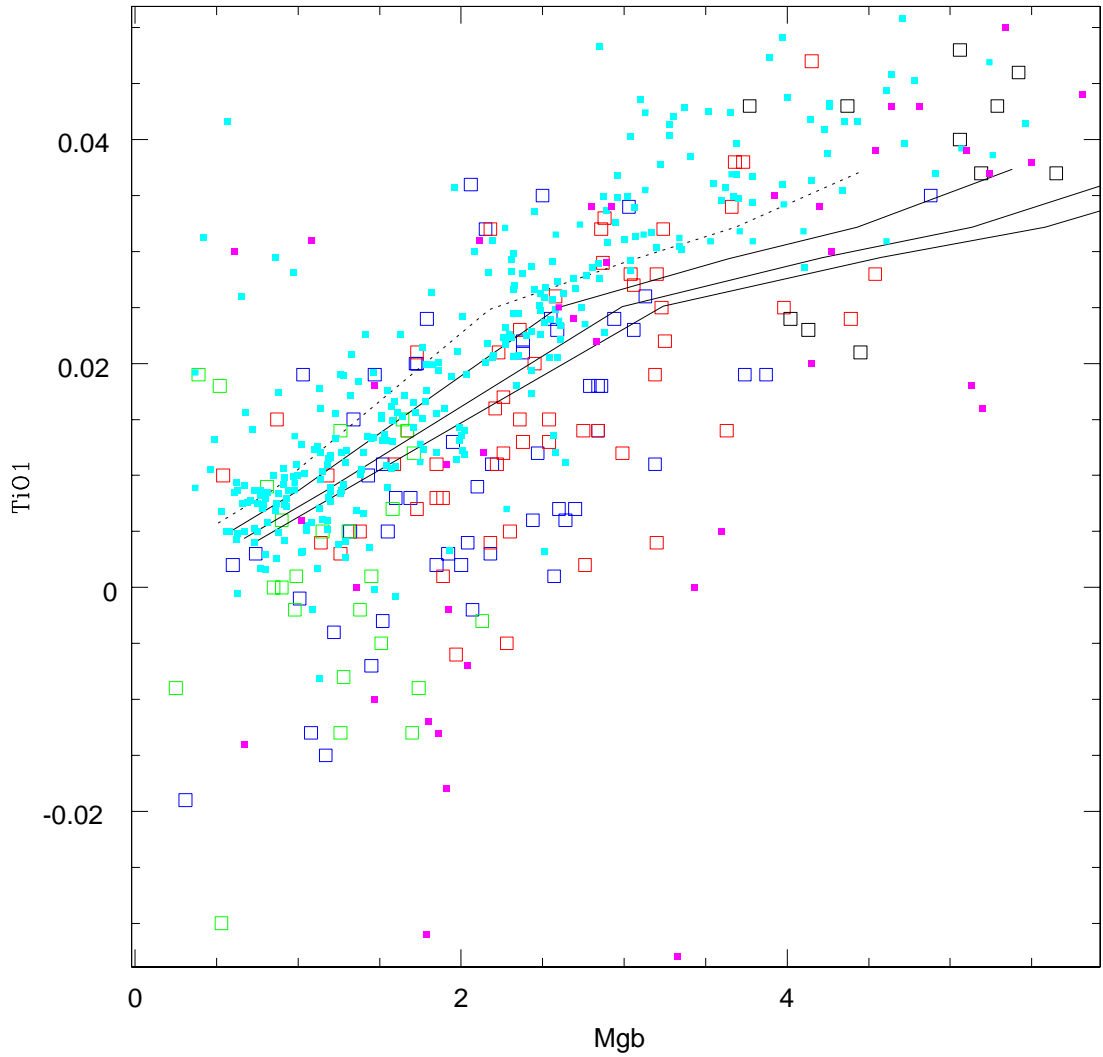}
\caption{TiO1 versus Mgb for G1 - G4 along with GCs of M31 and
NGC4472. The colours and symbols are the same as in Fig.4. TiO1 is
higher in the GCs of the spiral galaxy than in those of the
elliptical galaxies. } \label{Fig.5}
\end{center}
\end{figure}

\clearpage

\begin{figure}[h]
\begin{center}
\includegraphics[scale=0.75, angle=0]{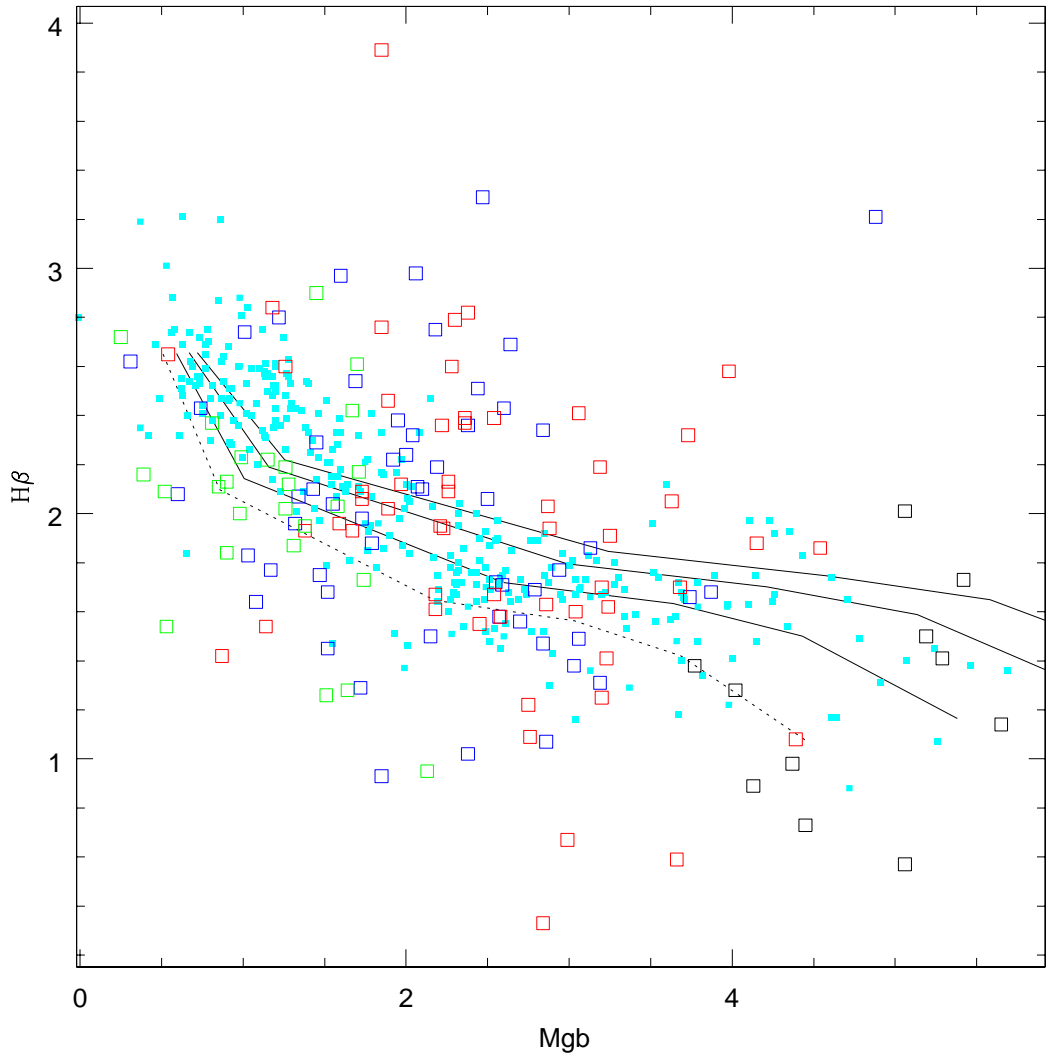}
\caption{$H{\beta}$ vesus Mgb for G1 - G4. The colours and symbols
are same as in Fig.4. The large spread in $H{\beta}$ in G1 - G4
can be due to a spread in $\alpha$/Fe and/or in color of the
Horizontal Branch. } \label{Fig.6}
\end{center}
\end{figure}

\clearpage

\begin{figure}[h]
\begin{center}
\includegraphics[scale=0.75, angle=0]{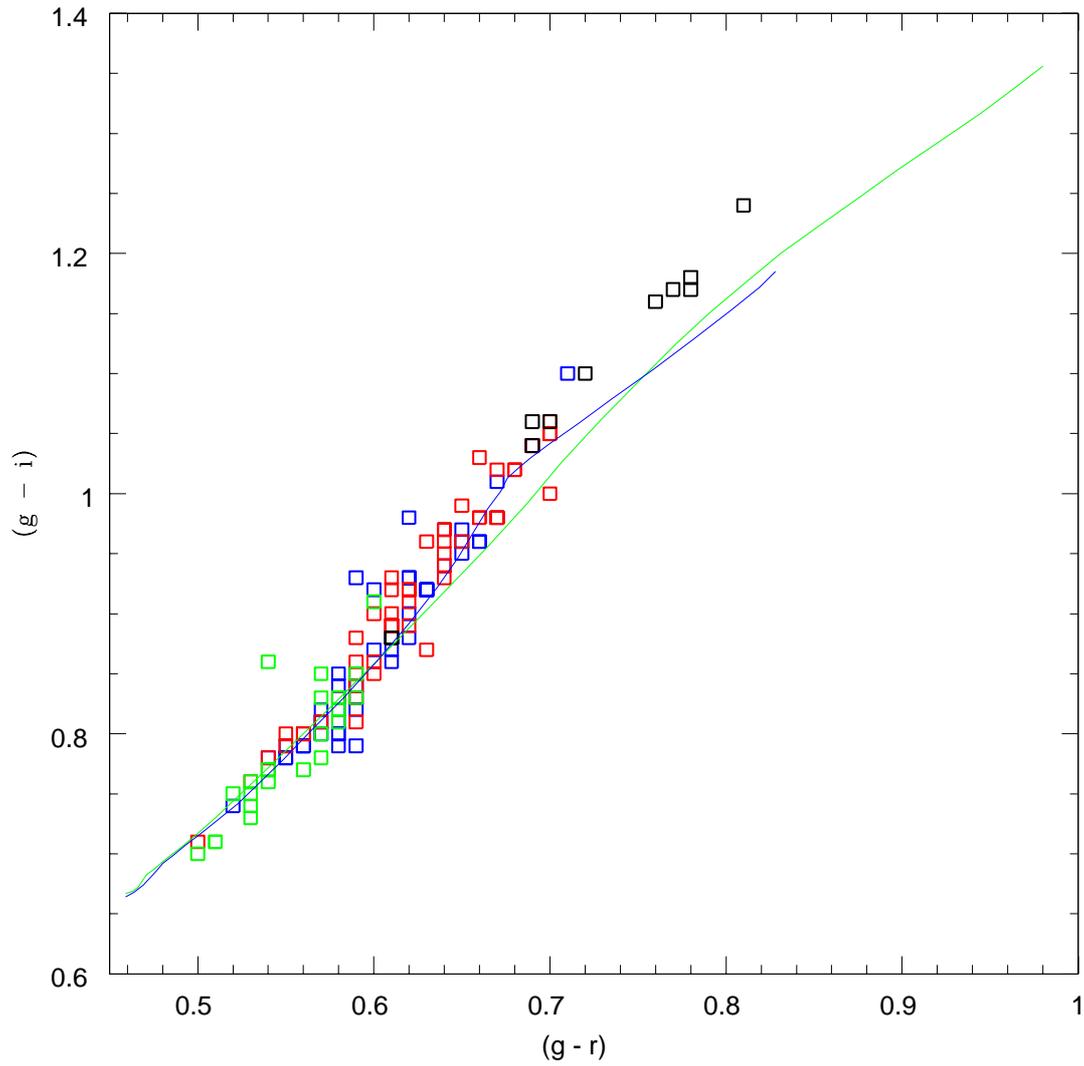}
\caption{(g-i) vs (g-r) for G1 - G4. The colours and symbols are
the same as in Fig.3. The green and blue solid lines are Yonsei
models for an age of 12Gyr, and for 0 and 70\% secondary stellar
populations respectively,
 } \label{Fig.7}
\end{center}
\end{figure}

\clearpage
\begin{figure}[h]
\begin{center}
\includegraphics[scale=0.75, angle=0]{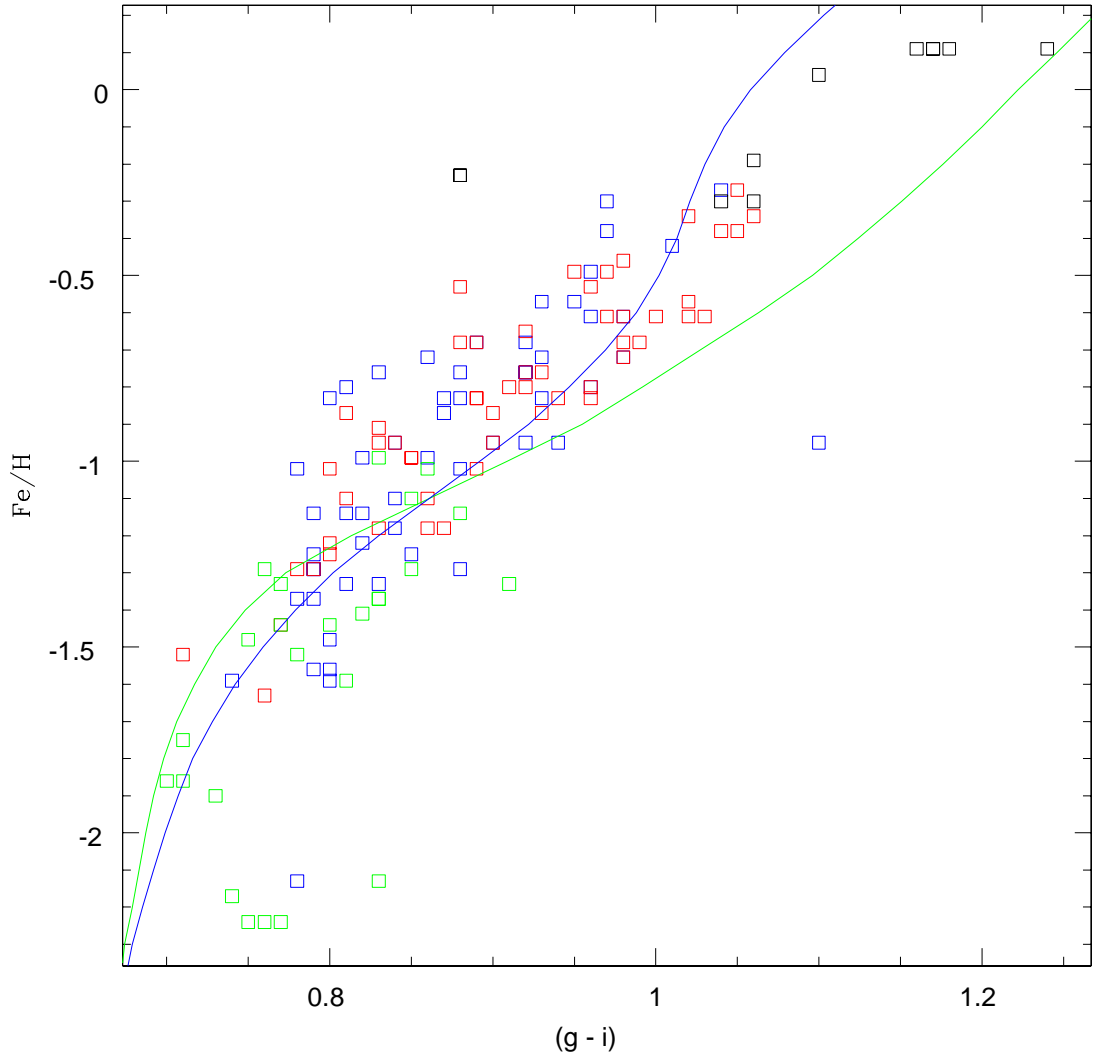}
\caption{[Fe/H] versus (g - i) for G1 - G4 . The colours and
symbols are same as that for Fig.3. The green and blue lines are
Yonsei stellar population models (Chung et al. 2013) for an age of
12 Gyr and 0 and 70 \% He enhanced secondary populations
repectively.  } \label{Fig.8}
\end{center}
\end{figure}
\clearpage

\begin{figure}[h]
\begin{center}
\includegraphics[scale=0.75, angle=0]{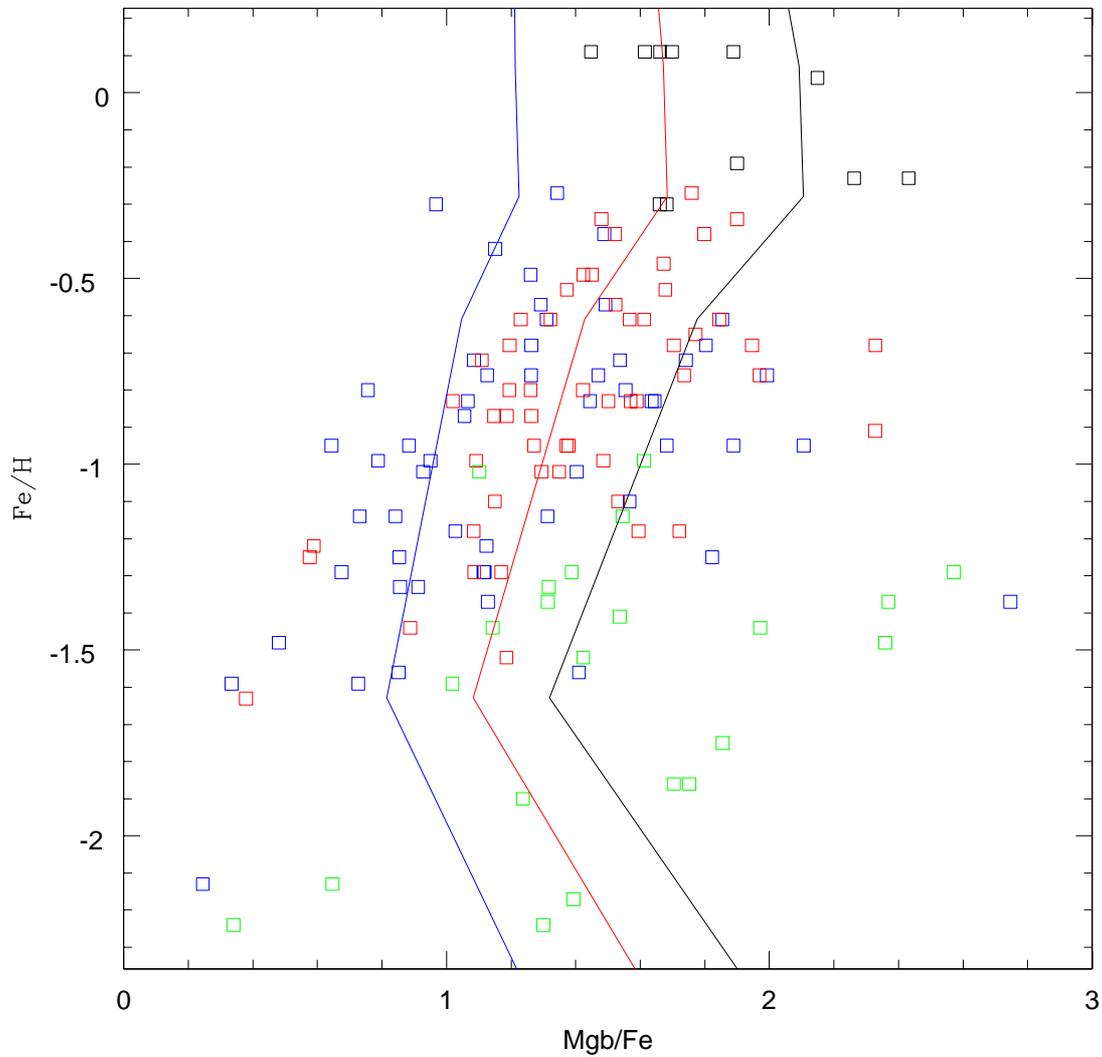}
\caption{Fe/H versus light-element abundance indicator Mgb/Fe for
G1 - G4. The colours and symbols are same as in Fig. 3. The blue,
red and black lines are stellar populations of Thomas et al.
(2011) for $\alpha$/Fe = 0.0, 0.3 and 0.5 respectively.
$\alpha$/Fe increases with metallicity in G1, G2 and G4, while in
G3 it has a peculiar behaviour.  } \label{Fig.9}
\end{center}
\end{figure}
\clearpage

\begin{figure}[h]
\begin{center}
\includegraphics[scale=0.75, angle=0]{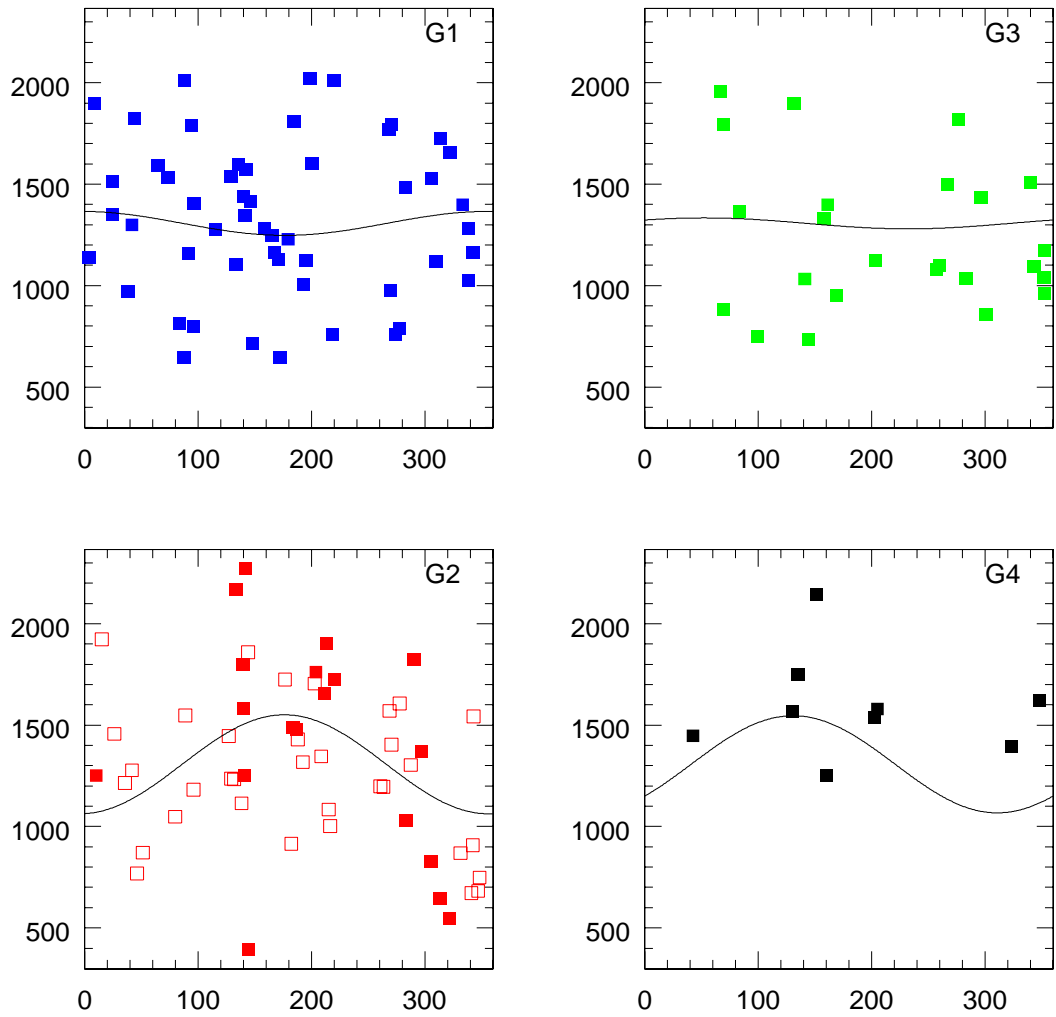}
\caption{$v_{rad}$ versus PA with fitted mean rotation curves for
G1 - G4. In the box for G2, the red open squares are for $R \leq
300 arcsec$ and the red full squares are for $R > 300 arcsec$. The
outer GCs of that group rotate faster than the inner GCs  }
\label{Fig.10}
\end{center}
\end{figure}

\clearpage
\begin{table}
\centering \caption{Discriminant analysis for K means
classification : G1, G2, G3, G4 are groups found by K means
Cluster Analysis and $G1^*$, $G2^*$, $G3^*$ and $G4^*$ are groups
found by Discriminant Analysis}
\begin{tabular}{c c c c c}
\hline \hline\\ DA groups & G1& G2& G3 &G4\\
$G1^*$ & 51 & 2& 3& 0\\
$G2^*$ & 2& 52& 1& 1\\
$G3^*$ & 2& 0& 23& 0\\
$G4^*$ & 0 & 0& 0& 10\\
Total& 55 & 54& 27 & 11\\
 \hline \hline
\end{tabular}
\end{table}


\begin{table}
\centering \caption{Mean values of various parameters of G1-G4
together with kinematics}
\begin{tabular}{c c c c c}
\hline \hline\\ Parameter & G1& G2& G3 &G4\\[0.5ex] \hline\\
No of GCs &55 &54 &27 &11\\
$H_\beta$ & 2.0271$\pm$0.0735 &1.9489$\pm$0.08 &2.0607$\pm$0.08
&1.238$\pm$0.13\\
Mg1& -0.00907$\pm$0.004 &-0.01204$\pm$0.004 &-0.0323$\pm$0.005
&0.1015$\pm$0.01\\
Mg2& 0.08618$\pm$0.007 &0.09270$\pm$0.007 &0.01981$\pm$0.009
&0.2781$\pm$0.02\\
Mgb& 2.123$\pm$0.115 &2.537$\pm$0.12 &1.1511$\pm$0.09
&4.765$\pm$0.19\\
Fe5270& 2.019$\pm$0.102 &1.9381$\pm$0.073 &0.711$\pm$0.13
&2.876$\pm$0.17\\
Fe5335& 1.5482$\pm$0.09 &1.6481$\pm$0.08 &0.5448$\pm$0.09
&2.405$\pm$0.24\\
NaD& 1.5122$\pm$0.09 &1.6646$\pm$0.001 &1.03$\pm$0.11
&5.125$\pm$0.52\\
TiO1& 0.01169$\pm$0.002 &0.0172$\pm$0.04 &0.00081$\pm$0.002
&0.03682$\pm$0.003\\
Fe/H& -0.9807$\pm$0.05 &-0.832$\pm$0.07 &-1.6289$\pm$0.08
&-0.06$\pm$0.06\\
Mgb/Fe&1.2318$\pm$ 0.0628&1.4184$\pm$0.0513&3.49$\pm$1.36&1.8543$\pm$0.0925\\
$i_{mag}$& 20.229$\pm$0.06 &20.077$\pm$0.006 &20.476$\pm$0.06
&20.002$\pm$0.13\\
g-r& 0.60291$\pm$0.005 &0.61833$\pm$0.01 &0.55037$\pm$0.006
&0.72$\pm$0.02\\
g-i& 0.8727$\pm$0.01 &0.9076$\pm$0.01 &0.7837$\pm$0.01
&1.0855$\pm$0.04\\
$R$& 147.73$\pm$8.36 &338.03$\pm$8.92 &234.2$\pm$17.9
&146.4$\pm$19.6\\
\hline
$v_{rot}$&30.92 $\pm$ 7.42 &232.96 $\pm$ 69.73 &40.93 $\pm$ 10.34 &249.02 $\pm$ 19.11 \\
$\psi_o$ & 92.58 $\pm $ 14.27& 265.38 $\pm $ 19.41& 167.20 $\pm $ 13.59 & 217.14 $\pm$ 50.65\\
\hline \hline
\end{tabular}
\end{table}

\clearpage

\section{Acknowledgements}

\noindent  We thank C. Chung for sending us tables for the Yonsei
Evolutionary Population Synthesis models. We are also thankful to
A.K.Chattopadhyay for useful discussions, and to Saptarshi Mondal
for helping in some calculations. One of the authors (Tanuka
Chattopadhyay) thanks DST, India for providing her a major
research project for the work.



\begin{thebibliography}{}


\bibitem [Albazzaz et al.(2004)] {alb04}  Albazzaz, H., Wang, X.Z.,
2004, Industrial \& Engineering Chemistry Research, 43(21), 6731.

\bibitem [Babu et al.(2009)] {bab09} Babu, G.J, Chattopadhyay,T., Chattopadhyay, A.K., Mondal, S., 2009,
ApJ,700,1768.

\bibitem [Bekki \& Chiba (2005)] {bek05} Bekki, K.,
Chiba,M., 2005, ApJ, 625, L107.

\bibitem [Betoule et al.(2013)] {bet13} Betoule,M., Marriner, J.; Regnault, N.; Cuillandre, J.-C.; Astier, P.;
Guy, J.; Balland, C.;  et al. 2013, A$\&$A, 552, A124.

\bibitem [Brodie \& Strader (2006)] {bro06}
Brodie,J.P., Strader,J., 2006, ARA\&A 44, 193.

\bibitem [Chattopadhyay et al.(2013)] {cha13}
Chattopadhyay, T., Karmakar, P., 2013a, New Astron. 22, 22.

\bibitem [Chattopadhyay et al. (2013)] {cha13}
Chattopadhyay, A. K., Mondal, S., Chattopadhyay, T., 2013b, CSDA,
57, 17

\bibitem [Chattopadhyay et al. (2009)] {cha09}
Chattopadhyay, A.K., Chattopadhyay, T., Davoust, E., Mondal, S.,
2009, ApJ, 705, 1533

\bibitem [Chattopadhyay et al.(2010)] {cha10}
Chattopadhyay, T., Sharina, M., Karmakar, P., 2010, ApJ, 724, 678

\bibitem [Chattopadhyay et al. (2012)] {cha12} Chattopadhyay, T., Sharina,
M., Davoust, E., De,T., Chattopadhyay, A.K., 2012, ApJ, 750, 91.

\bibitem [Chattopadhyay et al. (2007)] {cha07}Chattopadhyay, T., Chattopadhyay, A. 2007, A\&A, 472, 131

\bibitem [Chung et al. (2013)] {ch13} Chung, C., Lee, S.Y., Yoon, S.K, Lee, Y.W, 2013, ApJ,769, L3


\bibitem [Cohen \& Ryzhov (1997)] {coh97} Cohen, J.G.,
Ryzhov, A., 1997, ApJ, 486, 230

\bibitem [Cohen et al.(1998)] {coh98} Cohen, J.G.,
Blakeslee, J.P., Ryzhov.A., 1998, ApJ, 808, 539.

\bibitem [Cohen et al.(2003)] {coh03} Cohen, J.G.,
Blakeslee, J.P., C\^ot\'e, P., 2003, ApJ 592, 866

\bibitem [Colucci et al. (2012)] {co1a} Colucci, J.E., Bernstein, R.A., Cameron, S.A., McWilliam, A., 2012, ApJ, 746, 29

\bibitem [Colucci et al. (2012)] {co1b} Colucci, J.E., Bernstein, R.A., Cohen, J, 2012, Proc. XII International Symp. on Nuclei in the Cosmos, 2012, 99

\bibitem [Comon(1994)] {com94} Comon, P., 1994, Signal Processing, 36, 287.

\bibitem [Cote et al.(2001)] {cot01} C\^ot\'e, P., McLaughlin, D.E., et al, 2001, ApJ, 559, 828.

\bibitem [Eggen et al.(1962)] {egg62} Eggen, O.J., Lynden-Bell, D., Sandage, A. R.,
1962, ApJ 136, 748

\bibitem [Emselhem et al.(2007)] {ems07}
Emselhem, E., Cappellari, M., Krajnovic, D., et al, 2007, MNRAS,
379, 401

\bibitem [Fraix-Burnet et al.(2010)] {fra10} Fraix-Burnet, D.,
Chattopadhyay, T., Chattopadhyay, A.K., Davoust, E., 2010, MNRAS,
407, 2207

\bibitem [Fraix-Burnet et al.(2010)] {fra10} Fraix-Burnet, D., Davoust, E. and Charbonnel, C. 2009, MNRAS, 398, 1706.

\bibitem [Hanes et al.(2001)] {han01} Hanes, D.A., C\"ot\'e, P.,
Bridges, T.J., McLaughlin, D.E., Geisler, D., Harris, G.L.H.,
Hesser, J.E., Lee, M.G., 2001, ApJ 559, 812

\bibitem [Huchra \& Brodie(1987)] {huc87} Huchra,
J.P., Brodie, J.P., 1987, AJ, 93, 779.

\bibitem [Jesseit et al.(2009)] {jes09} Jesseit,
R., Cappellari, M., Naab, T., et al, 2009, MNRAS, 397, 1202.

\bibitem [Johnson \& Wichern (1998)] {joh98} Johnson, R. A., \& Wichern, D. W. 1998, Applied Multivariate
Statistical Analysis (4th ed.; Upper Saddle River: Prentice Hall)

\bibitem [Kaviraj et al. (2007)] {ka}
Kaviraj, S.; Sohn, S. T., O'Connell, R. W., Yoon, S.-J., Lee, Y.
W., Yi, S. K., 2007, MNRAS, 377, 987

\bibitem [Kotulla et al. (2009)] {ko} Kotulla, R., Fritze, U., Weilbacher, P., \& Anders, P.,
2009, MNRAS 396, 462

\bibitem [Krajnovic et al.(2008)] {kra08} Krajnovic, D., Bacon, R., Cappellari, M., Davies, R. L.,
de Zeeuw, P. T., Emsellem, E., Falcón-Barroso, J., et al. 2008,
MNRAS, 390. 93.

\bibitem [Kissler-Patig \& Gebhardt19(1998)] {kis98} Kissler-Patig,
M. \& Gebhardt, K. 1998, AJ, 116, 2237.

\bibitem [Lada \& Lada (2003)] {lad03} Lada, C.J.,
Lada, E.A., 2003, ARA\&A, 41, 57.

\bibitem [Lambert \& Gontier (2009)] {lam09}
Lambert, S.B., Gontier, A.-M., 2009, A\&A 493, 317.

\bibitem [Levenberg (1944)] {lev94}
Levenberg, K., 1944, Q Appl. Math., 2, 164.

\bibitem [Macqueen (1967)] {mac67}
Macqueen, J., 1967, in Proc 5th Berkeley Symp. Math Prob. Vol.1,
ed. L.M. Lecom and J. Neyman (Los Angeles Univ of California
press), 281.

\bibitem [Maratson et al.(2003)] {mar03}
Maratson, C., Greggiol, Renzini, A., et al, 2003, A$\&$A, 400,
823.

\bibitem [Marquardt (1963)] {mar63}
Marquardt, D.W., 1963, SIAMJ, Appl. Math., 11, 431.

\bibitem [Martell \& Grebel (2010)] {mar10} Martell,
S.L., and Grebel, E.K., 2010, A$\&$A, 519, 14.

\bibitem [\protect\citeauthoryear {Montes} {2014b}] {mon14} Montes,
M., Trujillo, I., et al, 2014b, MNRAS, 439, 990.

\bibitem [Montes et al.(2014a)] {mon14} Montes,
M., Trujillo, I., Prieto, M.A., Acosta-Pulido, J.A., 2014a, MNRAS,
439, 990.

\bibitem [Muratov et al. (2010)] {mur10} Muratov,
A.L. $\&$ Gnedin, O.Y., 2010, ApJ, 718, 1266.

\bibitem [Mould et al.(1987)] {mou87} Mould,
J.R., Oke, J.B., Nemec, J.M., 1987, AJ, 93, 53.

\bibitem [Mould et al.(1990)] {mou90} Mould,
J.R., Oke, J.B., De Zeeuw, P.T., Nemec, J.M., 1990, AJ, 99, 1823.

\bibitem [Park et al.(2012)] {par12}
Park, H.S., Lee, M.G., Hwang, H.S., Kim, S.C., Arimoto, N.,
Yamada, Y., Tamura, N., Onodera, M., 2012, ApJ 759, 116

\bibitem [Proctor et al.(2009)] {pro09} Proctor, R.N.,
Forbes, D.A., Romanwosky, A.J., et al, 2009, MNRAS, 398, 91.

\bibitem [Richtler et al.(2004)] {ric04}
Richtler, T., Dirsch, B., Gebhradt, K., Geisler, D., et al, 2004,
AJ, 127, 2094.

\bibitem [Schubert et al.(2010)] {sch10}
Schuberth Y., Richtler T., Hilker M., Dirsch B., Bassino L.P.,
Romanowsky A.J., Infante L., A\&A 2010, 513, A52

\bibitem [Schiavon et al.(2012)] {sch12} Schiavon, R. P., Caldwell, N., Morrison,
H., Harding, P., Courteau, S., MacArthur, L. A., Graves, G. J.
2012, AJ, 143, 14.

\bibitem [Shapiro et al.(1965)] {sha65} Shapiro, S.S., Wilk, M.B., 1965, Biometrika, 52 , 591.

\bibitem [Smith et al.(2000)] {smi00}
Smith, R. J., Lucey, J. R., Hudson, M. J., Schlegel, D. J.,
Davies, R. L. 2000, MNRAS, 313, 469

\bibitem [Sohn et al. (2006)] {so}
Sohn, S. T., O'Connell, R. W., Kundu, A., Landsman, W. B.,
Burstein, D., Bohlin, R. C., Frogel, J. A., Rose, J. A., 2006, AJ
131, 866

\bibitem [Strom et al.(1981)] {str81} Strom,
S.E., Strom, K.M., Wells, D.C., Forte, J.C., Smith, M.G., Harris,
W.E., 1981, ApJ, 245(5457), 416.

\bibitem [Strader et al.(2011)]  {str11} Strader, J., Romanowsky, A. J., Brodie, J. P., Spitler,
L. R., Beasley, M. A.; Arnold, J. A., Tamura, N.et al. 2011, ApJS,
197, 33.

\bibitem [Sugar \& James (2003)] {sug03} Sugar,
A.S., James, G.M., 2003, J.Am. Stat. Assoc., 98, 750.

\bibitem [Thomas et al.(2011)] {tho11}
Thomas D., Maraston C., Johansson, J., 2011, MNRAS 412, 2183

\bibitem [Woodley et al.(2007] {woo07} Woodley,
K.A., Harris, W.E., Beasley, M.A., et al, 2007, AJ, 134, 494.


\end{thebibliography}
\end{document}